
\input harvmac

\def\Titleh#1#2{\nopagenumbers\abstractfont\hsize=\hstitle\rightline{#1}%
\vskip .5in\centerline{\titlefont #2}\abstractfont\vskip .5in\pageno=0}

\def\CTPa{\it Center for Theoretical Physics, Department of Physics,
      Texas A\&M University}
\def\CTPb{\it College Station, TX 77843-4242, USA}
\def\HARCa{\it Astroparticle Physics Group,
Houston Advanced Research Center (HARC)}
\def\HARCb{\it The Woodlands, TX 77381, USA}

\def\CERN{\it CERN Theory Division, 1211 Geneva 23, Switzerland}
\def\ie{\hbox{\it i.e.}}     
\def\eg{\hbox{\it e.g.}}

\def\coeff#1#2{{\textstyle{#1\over #2}}}

\catcode`\@=11 

\def\lsim{\mathrel{\mathpalette\@versim<}}
\def\gsim{\mathrel{\mathpalette\@versim>}}
\def\@versim#1#2{\vcenter{\offinterlineskip
    \ialign{$\m@th#1\hfil##\hfil$\crcr#2\crcr\sim\crcr } }}
\def\boxit#1{\vbox{\hrule\hbox{\vrule\kern3pt
      \vbox{\kern3pt#1\kern3pt}\kern3pt\vrule}\hrule}}

\def\etal{{\it et. al.}}

\def\t1{{\tilde 1}}

\def\F{\widetilde F}

\def\JL{J. L. Lopez}
\def\DVN{D. V. Nanopoulos}
\def\AZ{A. Zichichi}
\def\HP{H. Pois}
\def\XW{X. Wang}

\def\GeV{\,{\rm GeV}}

\def\NPB#1#2#3{Nucl. Phys. B {\bf#1} (19#2) #3}
\def\PLB#1#2#3{Phys. Lett. B {\bf#1} (19#2) #3}

\def\PRD#1#2#3{Phys. Rev. D {\bf#1} (19#2) #3}

\def\PRT#1#2#3{Phys. Rep. {\bf#1} (19#2) #3}

\nref\AEN{I. Antoniadis, J. Ellis, and \DVN, \PLB{262}{91}{109}.}
\nref\Clavelli{L. Clavelli, \PRD{46}{92}{2112}; see also \PRD{45}{92}{3276}.}
\nref\CCYI{L. Clavelli, P. Coulter, and K. Yuan, \PRD{47}{93}{1973}.}
\nref\CCYII{L. Clavelli, \etal, \PLB{291}{92}{426};
L. Clavelli, P. Cox, and K. Yuan, University of Alabama preprint UAHEP9211.}
\nref\JK{M. Je\.zabek and J. K\"uhn, \PLB{301}{93}{121}.}
\nref\ENR{J. Ellis, \DVN, and D. Ross, CERN preprint CERN-TH.6824/93.}
\nref\Dickreview{For reviews see: R. Arnowitt and P. Nath, {\it Applied N=1
Supergravity} (World Scientific, Singapore 1983); H. P. Nilles,
\PRT{110}{84}{1}.}
\nref\LN{For a review see, A. B. Lahanas and D. V. Nanopoulos,
\PRT{145}{87}{1}.}
\nref\aspects{S. Kelley, \JL, \DVN, H. Pois, and K. Yuan, \NPB{398}{93}{3}.}
\nref\LNPWZh{\JL, \DVN, \HP, \XW, and \AZ, \PLB{306}{93}{73}.}
\nref\GH{See \eg, J. Gunion and H. Haber, \NPB{272}{86}{1}.}
\nref\HHG{{\it The Higgs Hunter's Guide}, J. Gunion, H. Haber, G. Kane, and
S. Dawson (Addison-Wesley, Redwood City 1990).}
\nref\ssb{L. Ib\`a\~nez and D. L\"ust, \NPB{382}{92}{305}; B de Carlos,
J. Casas, and C. Mu\~noz, CERN-TH.6436/92 and \PLB{299}{93}{234}; V.
Kaplunovsky and J. Louis, CERN-TH.6809/93.}
\nref\DM{\JL, \DVN, and K. Yuan, \NPB{370}{92}{445}; S. Kelley, \JL, \DVN, H.
Pois, and K. Yuan, \PRD{47}{93}{2461}.}
\nref\Valle{F. de Campos and J. Valle, FTUV/93-9 (February 1993).}

\Titleh{\vbox{\baselineskip12pt
\hbox{CERN-TH.6890/93}
\hbox{CTP--TAMU--25/93}
\hbox{ACT--07/93}}}
{\vbox{\centerline{LEP and Radiative Electroweak Breaking}
\centerline{Close the Light Gluino Window}}}
\centerline{JORGE~L.~LOPEZ$^{(a)(b)}$, D.~V.~NANOPOULOS$^{(a)(b)(c)}$, and
XU~WANG$^{(a)(b)}$,}
\smallskip
\centerline{$^{(a)}$\CTPa}
\centerline{\CTPb}
\centerline{$^{(b)}$\HARCa}
\centerline{\HARCb}
\centerline{$^{(c)}$\CERN}
\vskip .5in
\centerline{ABSTRACT}
We show that the LEP lower bound on the chargino mass, in conjunction with
the well motivated theoretical requirement of radiative electroweak symmetry
breaking, imply an upper bound on the lightest Higgs boson mass
($m_h\lsim62\GeV$) in a supersymmetry breaking scenario where the gluino mass
is a few GeV. Moreover, Higgs searches at LEP in the context of this model
require $m_h\gsim61\GeV$. The remaining experimentally allowed region in the
five-dimensional parameter space of this light gluino model is severely
fine-tuned (with $\tan\beta=1.88-1.89$ and $m_t=114.0-114.3\GeV$) and
cosmologically troublesome (with a neutralino relic abundance over {\it
five-hundred times larger than allowed}). Modest improvements in sensitivity in
LEP Higgs searches and Tevatron top-quark searches should soon exclude this
theoretically disfavored model completely.
\bigskip
\bigskip
\hbox{CERN-TH.6890/93}
\hbox{CTP--TAMU--25/93}
\hbox{ACT--07/93}
\Date{April, 1993}

There has been recent renewed interest on the viability of a narrow range of
presumably as yet experimentally unexplored light gluino masses ($m_{\tilde
g}\approx3-5\GeV$) \refs{\AEN,\Clavelli,\CCYI,\CCYII,\JK,\ENR}. This
possibility is motivated by the observation that such light colored particle
affects the running of the strong coupling and may make experimental
determinations of $\alpha_s$ at various low- and  high-energy scales to be in
better or worse agreement with each other \refs{\AEN,\Clavelli,\JK}. Recently,
\ENR\ a thorough re-examination of the compatibility of this scenario with
perturbative QCD data has yielded proposed experimental searches at
LEP and HERA which could let this matter to rest. In this paper we show that
well motivated theoretical assumptions in the context of unified supersymmetric
models, combined with current LEP data on chargino and Higgs searches, strongly
disfavor the light gluino scenario. In fact, the actual five-dimensional window
of parameter space in this class of light gluino models is almost completely
closed, except for some highly fine-tuned values of the parameters which are
unlikely to survive mild improvements in sensitivity in Higgs and top-quark
searches. Moreover, such values of the parameters imply a neutralino
cosmological relic density over {\it five-hundred times larger than
observationally allowed}.

We work in the context of unified supersymmetric models, where consistency
requires the introduction of supergravity to induce the low-energy
splittings between particle and sparticle masses \Dickreview. This procedure
takes the form of universal soft-supersymmetry breaking at the unification
scale ($M_U={\cal O}(10^{16}\GeV)$) and subsequent renormalization group
scaling of scalar and gaugino masses down to low energies. We also demand
radiative electroweak breaking of the electroweak symmetry \LN, which relates
the soft-supersymmetry breaking parameters to the Higgs mixing parameter $\mu$.
This class of unified supersymmetric models has been studied for over a decade
(for a recent reappraisal see \eg, Ref. \aspects). In this context all gaugino
mases, $m_{\tilde g}$ ($SU(3)_C$), $M_2$ ($SU(2)_L$), and $M_1$ ($U(1)_Y$), are
equal to $m_{1/2}$ at  $M_U$ and then scale at low energies as
\eqn\AI{m_{\tilde g}=m_{1/2}\alpha_s(M_Z)/\alpha_s(M_U)\approx2.9 m_{1/2},}
for $\alpha_s(M_Z)=0.120$, and
\eqn\AII{M_1={5\over3}\tan^2\theta_w M_2,\quad M_2\approx0.8m_{1/2}\approx 0.3
m_{\tilde g}.}
Thus, $m_{\tilde g}\sim3-5\GeV$ implies $m_{1/2}\sim1-1.7\GeV$ and $2M_1\approx
M_2\sim0.8-1.4\GeV$. These mass relations imply that the lightest eigenstate
of the $4\times4$ neutralino mass matrix is very nearly a pure photino with
mass $m_{\chi^0_1}\approx {8\over3}\sin^2\theta_wM_2\approx{1\over6}m_{\tilde
g}$. Furthermore, the lightest chargino mass is given by
\eqn\A{m^2_{\chi^\pm_1}=M^2_W+\coeff{1}{2}\mu^2
-\coeff{1}{2}\sqrt{\mu^4+4M^2_W\mu^2+4M^4_W\cos^22\beta},}
where $M_2/M_W\ll1$ has been used. It is not hard to see \CCYI\ that the LEP
lower bound $m_{\chi^\pm_1}>{1\over2}M_Z$ implies an absolute upper bound on
$\tan\beta$,
\eqn\B{|\cos2\beta|<|\cos2\beta|_{max}=1-{1\over4\cos^2\theta_w}=0.674\quad
\Rightarrow\quad \tan\beta\lsim2.27,}
and a $\tan\beta$-dependent upper bound on $|\mu|$,
\eqn\C{\mu^2<\mu^2_{max}=4M^2_W\cos^2\theta_w
(\cos^22\beta_{max}-\cos^22\beta)<(96\GeV)^2.}
Since radiative electroweak breaking requires $\tan\beta>1$ \LN, we obtain a
definite allowed interval in $\tan\beta$ to be explored. Note that for
$\tan\beta=1$, $|\mu_{max}|=96\GeV$ whereas for $\tan\beta=2.27$,
$|\mu_{max}|=0$.

The relatively low allowed values for $\tan\beta$ hint to the possibility of
a light Higgs boson in this scenario. In fact, for $\tan\beta=2.27$,
$m^{tree}_h<|\cos2\beta|M_Z\approx61.5\GeV$. Since it is known that an
improved experimental lower bound on $m_h$ ($m_h\gsim60\GeV$) applies to this
class of models \LNPWZh, it is important to determine the largest allowed
values of $m_h$ in the light gluino scenario. To this end we need to
investigate: (i) whether $\tan\beta=\tan\beta_{max}$ is actually allowed by any
other constraints on the model, and (ii) what is the magnitude of the one-loop
corrections to $m^{tree}_h$.

The radiative electroweak symmetry breaking constraint is enforced by requiring
the scalar potential to have a proper minimum at the electroweak scale. This
entails two constraint equations which one uses to determine the values of
$|\mu|$ and the universal bilinear soft-supersymmetry breaking mass $B$.
Using the tree-level Higgs potential, the $|\mu|$ equation is \aspects
\eqn\D{\mu^2=X_0 m^2_0+X_{1/2}m^2_{1/2}-\coeff{1}{2}M^2_Z,}
where $m_0$ is the universal soft-supersymmetry breaking scalar mass, and
$X_0,X_{1/2}$ are functions of $m_t$ and $\tan\beta$. In particular,
\eqn\E{X_0=-1+\coeff{3}{2}{\tan^2\beta-1\over\tan^2\beta+1}(m_t/192\GeV)^2.}
Since $m_{1/2}\sim1\GeV$, and $|\mu|<|\mu_{max}|<96\GeV$, Eq. \D\ shows that
$m_0$ must be bounded above by $m_0\lsim\sqrt{3/2X_0}M_Z$. This upper bound can
be very weak if $\tan\beta$ and $m_t$ are chosen such that $X_0$ nearly
vanishes. This leads to a fine-tuning situation \aspects\ discussed below.
Otherwise, an upper bound on $m_0$ and thus on the squark masses results, which
restricts the magnitude of the one-loop corrections to $m^{tree}_h$ since these
are proportional to $\ln(m^2_{\tilde q}/m^2_t)$. We note that our calculations
below use the one-loop effective scalar potential, where simple relations
as in Eqs. \D\ and \E\ are not obtainable. Nonetheless, the argument still
holds but with somewhat shifted values of the parameters.

A more ``efficient" way to maximize the value of $m_h$ is by considering the
largest possible $\tan\beta$ values. However, as shown in Eq. \C, these imply
very low values of $|\mu|$. In turn, the second-to-lightest neutralino will
have a significant higgsino component and will contribute to
$\Gamma(Z\to\chi^0_2\chi^0_2)$ more than the LEP data allow. (The lightest
neutralino has a very suppressed coupling to the $Z$ since it is nearly a pure
photino eigenstate.) Thus, $\tan\beta$ is not allowed to reach its otherwise
maximum possible values.

All the above remarks have been verified explicitly by a direct search of the
parameter space with $m_{1/2}=1\GeV$; $\tan\beta=1.25,1.50,1.75,2.00,2.25$;
$m_t=130,160\GeV$; $m_0=40\to200\GeV$; $A=0,\pm m_0$. Lower values of $m_0$
would cause the sneutrino mass to fall below the LEP lower bound
($m_{\tilde\nu}\gsim42\GeV$) and higher values push $|\mu|$ above
$|\mu_{max}|$. This explicit search yielded a set of allowed points in
parameter space with $m_h=16-46\GeV$ and $\tan\beta\le1.75$. Shortly we will
show that all these Higgs masses are already experimentally excluded.

As mentioned above, if in Eq. \D, $X_0\approx0$, the upper bound on $m_0$ can
be relaxed. As well, potentially allowed values of $\tan\beta$ between 1.75 and
2.00 would have been missed by our explicit search above. A more detailed
search of the parameter space geared at exploring the small-$X_0$ region and
the highest values of $\tan\beta$, shows that $\tan\beta$ values as high as
1.89 are actually allowed. However, only if $X_0\approx0$ can one obtain Higgs
masses which are substantially higher than the corresponding tree-level maximum
($\approx51\GeV$ for $\tan\beta=1.89$). Using the tree-level expression in Eq.
\E\ for $\tan\beta=1.88\,(1.89)$ one obtains $X_0=0$ for
$m_t=117.2\,(117.6)\GeV$. In actuality, using the one-loop effective potential,
the critical value of $m_t$ is $m_t=114.0\,(114.3)\GeV$. In this case we obtain
$m^{max}_h\approx62\GeV$, with $m_{\chi^\pm_1}\approx45-46\GeV$ and
$m_0\approx500\GeV$ also required. Values of $m_h>58\,(60)\GeV$ are obtained
for $\tan\beta>1.80\,(1.87)$ with the appropriately fine-tuned value of
$m_t=111\,(114)\GeV$. It is then crucial to determine whether these points in
parameter space are excluded by the LEP Higgs searches or not.

In Ref. \LNPWZh\ we described a procedure to obtain the appropriate
experimental lower bound on $m_h$ for a given point in parameter space in this
class of models, given the fact that the current standard model Higgs mass
bound is $m_H>61.6\GeV$. The condition to be satisfied by allowed points in
parameter space is
\eqn\F{f\cdot\sigma(m_h)_{susy}<\sigma(61.6)_{SM},}
where $f={\rm BR}(h\to2{\rm jets})_{susy}/{\rm BR}(H\to2{\rm jets})_{SM}$,
$\sigma_{susy}=\sigma(e^+e^-\to Z^*h)$, and $\sigma_{SM}=\sigma(e^+e^-\to
Z^*H)$. The computation of the $f$-factor includes all possible kinematically
allowed $h$ decay modes. Of particular concern are the following ones:
$h\to\chi^0_1\chi^0_1,\chi^0_1\chi^0_2,\chi^0_2\chi^0_2$. However, since
$\chi^0_1\approx\tilde\gamma$, the $h$-$\tilde\gamma$-$\tilde\gamma$ and
$h$-$\tilde\gamma$-$\chi^0_i$ couplings nearly vanish \GH. Moreover, we
find $m_{\chi^0_2}>36\GeV$ making these decay channels inconsequential for
$m_h\lsim62\GeV$, and therefore $f\approx1$ \LNPWZh. The cross sections in Eq.
\F\ differ simply by a coupling factor $\sin^2(\alpha-\beta)$ and the Higgs
mass used which enters through a function $P$. With these substitutions we get
$f\cdot\sin^2(\alpha-\beta)<P(61.6/M_Z)/P(m_h/M_Z)$
with \HHG
\eqn\G{P(y)={3y(y^4-8y+20)\over\sqrt{4-y^2}}\cos^{-1}
\left(y(3-y^2)\over2\right)-3(y^4-6y^2+4)\ln
y-\coeff{1}{2}(1-y^2)(2y^4-13y+47).}
This analysis shows that only those points above with $m_h\gsim61\GeV$ would
still be experimentally allowed, \ie, with the fine-tuned values
$\tan\beta=1.88-1.89$, $m_t=114.0-114.3\GeV$, $m_{\chi^\pm_1}\approx45-46\GeV$,
and $m_0=480-540\GeV$. We note that a mild improvement in sensitivity in LEP
Higgs searches and Tevatron top-quark searches are likely to exclude this
remaning region of parameter space. One should bear in mind that the
fine-tuning required to get these allowed points would need
to be performed at each order in perturbation theory.

We expect the above situation to persist for a range of light
gluino masses ($m_{\tilde g}\sim m_{\tilde\gamma}\ll M_W$), as long as the
gaugino masses are related to each other by ${\cal O}(1)$ coefficients as
assumed here. We also note that the class of supersymmetry breaking scenarios
that is actually being studied here ($m_{1/2}/m_0\sim1/100$) is also obtained
in recent string-inspired model calculations \ssb, except that such low values
of $m_{1/2}$ have not been considered so far.

It is interesting to note that other phenomenological constraints may also
lead to stringent constraints on this scenario. In fact, if the lightest
neutralino (the photino) is stable, as it would be in the usual $R$-parity
conserving models, then its cosmological relic abundance would be quite large
due to the inefficiency of the possible annihilation channels. Following the
methods of Ref. \DM, we obtain $\Omega_\chi h^2_0\gsim33$ in general,
and $\Omega_\chi h^2_0\gsim500$ for $m_h\gsim61\GeV$. Therefore, the remaining
fine-tuned region of parameter space is in gross conflict with cosmological
observations (which require $\Omega_\chi h^2_0<1$) and could only be made
cosmologically acceptable in a model with suitable $R$-parity breaking (see
\eg, Ref. \Valle). We also note that future improvements in the determination
of the $W$-width at the Tevatron may be sensitive enough to the
$W^\pm\to\chi^\pm_1\tilde\gamma$ contribution, further constraining this
scenario.

We conclude that the theoretical requirement of radiative electroweak symmetry
breaking in the light gluino scenario entails a rather light Higgs boson mass
range ($m_h\lsim62\GeV$) which is almost completely in conflict with LEP
search limits,\foot{{\it c.f.} Ref. \Valle\ where this constraint was not
applied and the light gluino scenario in supergravity models was deemed
viable.} which in this model imply $m_h\gsim61\GeV$. Moreover, the very small
experimentally allowed region in parameter space is cosmologically troublesome
and severely fine-tuned with $\tan\beta=1.88-1.89$ and $m_t=114.0-114.3\GeV$.
It should not be long before further searches at LEP ($m_h$) and the Tevatron
($m_t$) close the light gluino window completely in the well motivated class
of supergravity models we consider.

\bigskip
\bigskip
\bigskip
\bigskip
\noindent{\it Acknowledgments}: This work has been supported in part by DOE
grant DE-FG05-91-ER-40633. The work of J.L. has been supported by an SSC
Fellowship. The work of  D.V.N. has been supported in part by a grant from
Conoco Inc. The work of X.W. has been supported by a World-Laboratory
Fellowship. J.L. would like to thank Kajia Yuan, Gye Park, and Heath Pois
for useful discussions.
\listrefs
\bye